# Second-Nearest-Neighbor Correlations from Connection of

# Atomic Packing Motifs in Metallic Glasses and Liquids

Jun Ding[1], Evan Ma[2], Mark Asta[1,3]*, Robert O. Ritchie[1,3]*

[1] Materials Sciences Division, Lawrence Berkeley National Laboratory, Berkeley, California 94720, USA

[2] Department of Materials Science and Engineering, Johns Hopkins University, Baltimore, Maryland 21218, USA.

[3] Department of Materials Science and Engineering, University of California, Berkeley, California 94720, USA.

mdasta@berkeley.edu;  roritchie@lbl.gov

## Abstract

Using molecular dynamics simulations, we have studied the atomic correlations characterizing the second peak in the radial distribution function (RDF) of metallic glasses and liquids. The analysis was conducted from the perspective of different connection schemes of atomic packing motifs, based on the number of shared atoms between two linked coordination polyhedra. The results demonstrate that the cluster connections by face-sharing, specifically with three common atoms, are most favored when transitioning from the liquid to glassy state, and exhibit the stiffest elastic response during shear deformation. These properties of the connections and the resultant atomic correlations are generally the same for different types of packing motifs in different alloys. Splitting of the second RDF peak was observed for the inherent structure of the equilibrium liquid, originating solely from cluster connections; this trait can then be inherited in the metallic glass formed via subsequent quenching of the parent liquid through the glass transition, in the absence of any additional type of local structural order. Increasing ordering and cluster connection during cooling, however, may tune the position and intensity of the split peaks.

## Introduction

Metallic glasses (MGs) were first discovered some five decades ago but are still of significant current interest because of their unique structure and properties [1-4]. Indeed, many fundamental materials science issues remain unresolved for MGs, as well as for supercooled liquids (SLs) which are their parent phase above the glass transition temperature [1-6]. However, compared to crystalline materials, the lack of long-range translational order presents inherent challenges to characterizing the atomic-level structure, and to discerning the salient structure-property relationships in amorphous alloys [7,8]. These issues involving the atomic-level structure in MGs and SLs have been under extensive study in recent years [7-15]. Notably, their short-range order (SRO) has been characterized in terms of atomic packing motifs. These motifs are the common coordination polyhedra in each MG (each coordination polyhedron is for an atom at center with surrounding nearest neighbors, NNs). For example, some MGs are characterized by full icosahedra as the dominant motif, which are coordination polyhedra with Voronoi index <0, 0, 12, 0> and five-fold bonds only. In these alloys a variety of thermodynamic, kinetic and mechanical properties have been correlated with the degree of icosahedral SRO [16-19]. In general, the SRO has a diverse range in terms of the preferable motifs, as summarized for different MGs in Refs. 7 and 8.

At length scales longer than that typically described by this SRO, *i.e.*, beyond the first NNs corresponding to the first peak in the radial distribution function (RDF), characterizing the atomic structure of these materials becomes even more complicated and challenging [9,10,20-30]. For example, efficient packing of quasi-equivalent "clusters" [9, 10] (the motifs) has been proposed, where the packing of the polyhedra in three-dimensional space is pictured to follow an icosahedral or face-centered cubic (F.C.C) pattern. There is also the notion of possible self-similar packing of atomic clusters with the characteristics of a fractal network of dimension 2.31 or 2.50 [20,21]. Additionally, the concept of spherical-periodic order, derived from the resonance between static order and the electronic system, was modified to involve additional local crystal-like translational symmetry to describe atomic order up to the long-range scale [22-24,27]. However, before one can establish the nature of extended order (such as those postulated in these models, which most likely vary from one alloy system to another), a useful step is to first understand the atomic correlations with atoms in the second nearest neighbor shell. These latter correlations are reflected by the second peak in the RDFs of MGs and their parent SLs. As the distances characteristic of the second peak are just beyond the short-range scale, *i.e.*, the packing of atoms in the NN shell constituting the motif/cluster above, the second-NN correlations can be a useful starting point for the characterization of medium-range order.

A commonly used method to illustrate how one atom correlates with atoms in its second nearest-neighbor shell is using an analysis of connection schemes of the coordination polyhedra, where previous work has shown that the coordination polyhedra can connect with each other by sharing one, two, three or four atoms [31-36]. As such, the pair correlations giving rise to the second peak, and its splitting in many observations, may have a universal origin in the specific ways each motif can connect with the next, see for example, Ref. 35. The purpose of this article is to perform a systematic analysis of such connection schemes across a broader range of MG systems than has been considered previously. We will address questions including: (i) do cluster connection schemes vary across different MG systems with differing compositions and SRO (NN packing motifs); ii) how cluster connection schemes evolve as a function of temperature during cooling, in particular the difference between the liquid state and the glass; (iii) can the various connection schemes account for the split second RDF peak, across the board for different systems with differing packing motifs, and for the same system in the MG versus SL state; and (iv) how the different cluster connection schemes affect mechanical performance, *i.e.*, which cluster connections are stiffer or more flexible.

To address these issues, we have conducted a systematic study using molecular dynamics (MD) simulations [37] of a number of representative model metallic-glass systems with different constituents and prepared at different cooling rates (see Methods). These systems were modeled by embedded-atom-method potentials optimized for the following MG systems: $Cu_{64}Zr_{36}$, $Ni_{80}P_{20}$, $Al_{90}La_{10}$, $Mg_{65}Cu_{25}Y_{10}$ and $Zr_{46}Cu_{46}Al_8$ [34,38,39] (Table I). The SRO motifs in these samples have been characterized before: the topological packing of NN atoms (*i.e.*, within the first peak of the RDF) has been well documented [7].

**Results and Discussions**

**General properties of polyhedra connections**

To illustrate how the second-NN pair correlation distance is related to the cluster connection, Fig. 1(a) shows schematically two representative atoms that are second nearest neighbors. Each of these two atoms is of course the center of its own coordination polyhedron (cluster) [31-36]. The two clusters are represented by the two color-shaded regions in Fig. 1(a). They are connected together, by the atoms at the locations where the two clusters overlap (the shared atoms). For any arbitrary reference atom, its second NN shell can be pictured as composed of atoms each at the center of a cluster connected to that of the reference atom (see the example depicted in the inset in Fig. 1(a)). For all the atoms in the second NN shell, their spatial correlations with the reference atom superimpose into the second peak in the RDF, $g(r)$, as indicated in Fig. 1(a).

In Fig. 1(a) the RDF shown by the thick solid line is for a Ta equilibrium liquid at 3300 K (following the same simulation procedure in [40]), while the thinner cyan line reflects the corresponding inherent structure obtained by conjugate-gardient energy minimization (see Methods) to remove the vibrational thermal contributions. The inherent structure of a liquid, with these vibrational contributions excluded, represents the local minimum of the potential energy basin the liquid is in [5,6,41] and has been widely utilized to study the liquid structure [42-45]. The second peak in $g(r)$ of the inherent structure of Ta liquids is split, similar to the Ta glass (see blue dashed line in Fig. 1(a)) obtained by quenching the same liquid at $10^{13} \, K/s$ to room temperature. Such peak splitting has been observed in numerous amorphous metals and alloys, see for example, Refs. [22,25,31-36], and will be discussed in more detail later.

The cluster connection can have multiple possible schemes, as shown schematically in Fig. 1(b). Here the neighboring polyhedra share one, two, three and four atoms, respectively, which are denoted hereafter as *1-atom*, *2-atom*, *3-atom*, and *4-atom* connections, respectively. The first three categories refer to the connections by sharing a vertex, an edge and a face of polyhedra, while the last category (*i.e.*, *4-atom*) refers to sharing distorted quadrilateral or squashed tetrahedra (*i.e.*, with 4 atoms almost in the same plane). The latter category is different from the previous definition of interpenetrating polyhedra [16,34], where the two central atoms inside are nearest neighbor atoms instead of second nearest neighbors, which is the focus of this paper. The cluster connections beyond *4-atom* (sharing more than 4 atoms) are neglected due to their very low fraction (e.g., < 0.018 per atom in sample #8). Each of these different connection schemes in Fig. 1(b) leads to a different most-probable distance between the two center atoms, thus giving rise to peaks at different correlation distances in the RDF. In other words, for any given MG, the broad second peak in the RDF is a result of the superimposition of the contributions from the four connection schemes and would likely show sub-peaks.

In Fig. 2, the $g(r)$ for $Ni_{80}P_{20}$ (sample #5) and $Zr_{46}Cu_{46}Al_8$ (sample #8) MGs at 300 K are evaluated up to large atomic separations (20Å). The red arrows in Fig. 2(a) indicate the splitting of the second peak for $Ni_{80}P_{20}$, similar to that observed in Fig. 1(a) and in previous experiments and simulations in the literature [22,25,31-36]. Note that not all MGs exhibit split second peaks, e.g., see the $Zr_{46}Cu_{46}Al_8$ case in Fig. 2(b) and further discussions below.

The decomposed components of the RDFs, specifically for the NN atoms and four cluster connection schemes for atoms in the second-NN shell, can be defined as:

$$g_{\alpha,\text{decomp}}(r) = \frac{V}{4\pi r^2 N_\alpha N_\alpha} \sum \sum \delta \left( \left| r_{ij} - r \right| \right) \qquad (i, j \in \alpha) \qquad (1)$$

where $i$ and $j$ atoms are linked by $\alpha$ type of connection (NN, or one of the four atomic cluster connection schemes for the second NNs), $r_{ij}$ is their interatomic distance; $N_\alpha$ are the number of atoms with $\alpha$ type of connection; $V$ is the volume of the entire sample. For comparison, the decomposed RDF components for each of the connection schemes for $Ni_{80}P_{20}$ and $Zr_{46}Cu_{46}Al_8$ MGs are also plotted in Fig. 2. These decomposed $g(r)$ curves resemble those reported previously [33], but with a stronger intensity for *1-atom* and *3-atom* and a weaker intensity for *2-atom* and *4-atom* connections.

From the geometry seen in Fig. 1(b) for cluster connection schemes corresponding to *1-atom*, *2-atom*, *3-atom*, and *4-atom*, the most-probable distance ($R_2^*$) between the two second-NN atoms (the centers of the two connected coordination polyhedra) can be calculated to be [31] $2R_1$, $\sqrt{3}R_1$, $2\sqrt{\dfrac{2}{3}}R_1, \sqrt{2}R_1$, respectively, where $R_1$ is the average bond length. These are therefore the predicted second peak positions from the decomposed $g(r)$. To make a comparison between the MD simulations and these calculated $R_2^*$, for each of the four types of cluster connections we evaluated the second-NN distance (peak position), averaged over all the partial RDFs for each MD glass sample, for each species, using $\overline{R}_2^* = \dfrac{\int r \cdot g_{\alpha,\text{decomp}}(r)\,dr}{\int g_{\alpha,\text{decomp}}(r)\,dr}$. The results are plotted in Fig. 3(a), where each data point represents one species for a sample listed in Table I, and the solid lines represent the $R_2^*$ predictions. As can be seen in Fig. 3(a), the MD simulation results closely match the calculated $R_2^*$. This supports the notion that the atomic cluster connection is primarily responsible for the second-peak locations in the RDF.

Another important issue is if and how the cluster connections depend on the NN coordination number (CN). The CN varies with the local SRO, reflecting different atomic size ratio and cluster topological order for each amorphous system. In other words, the packing motif is different from alloy to alloy. This issue is examined in Fig. 3(b), where we plot the average number of cluster connections against the average CN surrounding each species, in various alloys (Table I) at the corresponding experimental liquidus temperatures ($T_l$) [46]. Although the experimental value of $T_l$ may not correspond exactly with the liquidus predicted by the EAM potential, we have also tested the temperature range between ($T_l$-100) K and ($T_l$+100) K, and the results in Fig. 3(b) are largely unaffected. Analysis at the liquidus temperatures undertaken here avoids the complexity associated with cluster connection development by structural ordering during cooling through the glass transition (illustrated in detail below in Fig. 4).

We observe that the dependence of *1-atom* and *2-atom* cluster connections on CN is weak, while the *3-atom* and *4-atom* cluster connections scale almost linearly with CN (apart from small fluctuations about linear behavior), as seen in Fig. 3(b). In other words, at the liquidus temperature, the larger the coordination number of a central atom, the more *3-atom* and *4-atom* cluster connections exist, while the number of *1-atom* and *2-atom* remain essentially unchanged. This observation likely results from the closer distance between the first nearest-neighbor shell and *3-atom* or *4-atom* connected clusters, which implies that the CN (and hence different motif) exerts an influence on the number of the clusters connected. Nevertheless, in all cases the connection schemes and associated characteristic $R_2^*$ values are universally the same. This is not surprising; as all the characteristic motifs tend towards polytetrahedral packing inside [7,8], for all MGs the cluster connection ultimately is the connection of tetrahedral units.

It should be noted that all the data in Fig. 3(b) were evaluated within equilibrium liquids rather than in the glassy state. The structural ordering during supercooling, especially close to the glass transition, will alter the preference for certain connection schemes, in particular of the *2-atom* and *3-atom* clusters, as described in the following section; the correlations in Fig. 3(b) in liquids will therefore not be necessarily the same for the glassy state. More discussion of the difference follows in the next section.

**Influence of structural ordering during cooling on cluster connections**

In Fig. 4(a), we show the cluster connection number per atom for the four schemes (*1-atom, 2-atom, 3-atom* and *4-atom*) at room temperature (300 K) for sample #1 to 8 in Table I. The number of *1-atom* and *4-atom* cluster connections are the highest and lowest respectively among all the MG samples studied in this work. The relative fractions of the four connection schemes obviously determine the make-up (constitution ratio) of the second peak in RDF, affecting its shape and sometimes causing its splitting (see next section). Meanwhile, in Fig. 4(b) the development of cluster connections (*1-atom to 4-atom*) among all the samples listed in Table I are illustrated. Specifically, each data point plotted is defined as the difference between the glassy state (at 300 K) and the liquid state (at the liquidus temperature). The results in Fig. 4(b) demonstrate firstly that the number of *2-atom*, *3-atom* and *4-atom* connections in an alloy changes as one cools from the equilibrium liquid into a glassy state with increased structural ordering. Specifically, the number of *3-atom* connections increases while the number of *2-atom* and *4-atom* connections is reduced in the glassy state relative to the liquid, similar with other observations [16,33,38]. In contrast, the average number of *1-atom* connections per atom remains almost unchanged when going from the equilibrium liquid to the glassy state. Secondly, the development of *2-atom*, *3-atom* and *4-atom* cluster connections depends on the cooling rate, as illustrated for the $Cu_{64}Zr_{36}$ subjected to different quenching procedures (marked in Fig. 4). Specifically, a sample experiencing a slower cooling rate undergoes more structural ordering, exhibits a more pronounced increase in *3-atom* connections and a decrease of the *2-atom* and *4-atom* connections upon cooling into a glass. Thirdly, examination of the total number of cluster connections in the equilibrium liquids and glassy states in Fig. 4(b) indicates that the increase in *3-atom* connections is roughly compensated by a decrease in the number of *2-atom* and *4-atom* connections such that the total number of connections remains essentially unchanged. What this implies is that the *3-atom* cluster connections are the most favored; their number is increased with structural ordering during cooling through the glass transition at the expense of two of the other cluster connections, specifically the *2-atom* and *4-atom* connections. In other words, two neighboring coordination polyhedra prefer to link together via face sharing rather than edge or squashed-tetrahedra sharing. This can be regarded as a characteristic structural feature of atomic order, for correlations with the second nearest-neighbor shell in amorphous alloys. The increased fraction of *3-atom* connections leads to a higher intensity at first sub-peak, which is indeed observed in the RDFs of all the samples we studied.

**Splitting of second peak in the radial distribution functions**

The split second peak in the RDF and structure factor is often the most eminent observation for some MGs, see Fig. 2(a) and Ref. [22,25,31-36]. However, this is not a universal phenomenon for all MGs, e.g., $Zr_{46}Cu_{46}Al_8$ samples in Fig. 2(b) lack the splitting in the second peak in the RDF (also see Ref. [49]). The origin of that split second peak for MGs has been the subject of some debate [22,25,33,47,48]. Since the splitting is not observed in the data for liquids, most explanations attribute the phenomenon to structural ordering during the transition from the liquid to the glassy state (the SRO develops at increasing rate deep inside the supercooled liquid regime before the glass transition [7]). Various mechanisms have been proposed to account for the splitting, such as the intensified icosahedral order [13], appearance of local translational symmetry [22,47], the enhanced unevenness of the connection scheme of the atomic cluster [33,35], or "Bergman triacontahedron" packing [25] during the glass transition. These explanations all appear to be self-consistent, but there still remains the fundamental question as to whether the splitting of the second peak has to originate intrinsically from the structural ordering near the glass transition. The answer is negative: earlier in Fig. 1(a), we have examined both instantaneous and inherent structure (IS) for an equilibrium Ta liquid at 3300 K, and we pointed out that their RDFs are significantly different as the second peak in $g(r)$ is already split for the inherent structure.

In other words, this splitting feature in the RDF appears to be intrinsic even for an equilibrium liquid, where extended icosahedral order or crystalline topological order is absent. This supports the proposition that its origin is cluster connection schemes, because the equilibrium liquid already has a tendency to develop certain type of preferable coordination polyhedra, which connect via the four types of connection schemes. In the instantaneous liquid, with deviations away from the inherent structure and smearing by thermal vibration, the splitting feature is not observed at high temperatures. One thus concludes that the splitting second peak in $g(r)$ for MGs can be inherited from the inherent structure of liquids, and not fundamentally determined by the appearance (or not) of new local structure order developed towards glass transition [22,25,33,47]. As illustrated in Fig. 2 for the four cluster connection schemes, the contributions from *1-atom* and *3-atom* connections are much stronger in intensity while the *2-atom* and *4-atom* connections are weaker. Their uneven contributions can cause the splitting of second peak in $g(r)$, as shown in Fig. 2(a) and discussed in Ref. [33,35]. Usually, the splitting of second peak is more pronounced for monoatomic MGs [27], or when the system contains elements of similar atomic sizes [27], or at low temperatures. Conversely, when there are multiple constituents, large atomic size difference, and strong chemical order or vibrational contributions, the $g(r)$ decomposed for each of the cluster connection schemes would get broadened, and their superimposition tends to smear out the split second peak in $g(r)$ (see Fig. 1(a) and Fig.2(b)).

However, the structural ordering during cooling through the glass transition does also have influence on the second peak in $g(r)$, in that the enhancement of *3-atom* connections at the expense of both *2-atom* and *4-atom* connections (see Fig. 4) is expected to cause a shift and intensity changes of the sub-peaks. For instance, Fig. 1(a) compares the $g(r)$ of the inherent structure of Ta liquid, and that of Ta MG; in the latter the first sub-peak is more pronounced.

**Cluster connection dependence of elastic deformation**

Unlike crystalline metals, the elastic deformation of MGs is intrinsically inhomogeneous due to the wide variation in the local structural arrangements [28,29,50-53]. Consequently, it is interesting to examine the elastic response with respect to different atomic cluster connections within amorphous solids. Here we use MD simulations to examine the athermal quasi-static shear (AQS) [54] deformation of samples #1 to 8 in Table I in the nominal elastic regime; we further calculate elastic strains generated between connected coordination polyhedra (calculations are described in the Methods section). In Fig. 5, we plot the average elastic shear strain for each group of cluster connection schemes (colored arrows) in comparison to the imposed macroscopic shear strain; the dashed line in the figure represents where cluster strain equals the imposed macroscopic strain. We observe that i) the elastic strain experienced by clusters connected via the *1-atom* classification is almost equivalent to that of the macroscopic deformation; ii) the elastic responses of clusters with both the *2-atom* and *4-atom* connections behave in a more flexible manner (*i.e.*, they show deformation larger than the macroscopic strain) while clusters with *3-atom* connections are the stiffest (*i.e.*, they show the smallest local shear strain). Sharing of triangulated faces between tetrahedra is likely to result in higher energy barrier $W$ of basins in the potential energy landscape, which is known to increase shear modulus $G$ [55,56].

Interestingly, the variations in elastic deformation for clusters with the different connections correlate with the observed evolution in the population of these different connection schemes upon cooling. Specifically, as discussed above, the fraction of *3-atom* cluster connections (stiffest elastic response among atomic cluster connections) grows during the structural ordering, while the number of *2-atom* and *4-atom* connections (more flexible elastic response) is reduced. Meanwhile, *1-atom* cluster connections, which exhibit the insensitive dependence on structure ordering, have elastic response equivalent to the macroscopic deformation.

**Methods**

**Sample preparation by MD simulation:** Molecular dynamics simulations using LAMMPS package [57] were implemented to study eight metallic glass models, including the metallic glasses $Cu_{64}Zr_{36}$, $Ni_{80}P_{20}$, $Al_{90}La_{10}$, $Mg_{65}Cu_{25}Y_{10}$ and $Zr_{46}Cu_{46}Al_8$ (Table 1), with the optimized embedded atom method (EAM) potential, adopted from [34,38,39]. Each model contained 128,000 atoms with a simulation box length in excess of 10 nm, *i.e.,* large enough to overcome possible issues from periodic boundary conditions for longer length-scale order in metallic glasses and liquids. The liquids for the MD samples were equilibrated for sufficient times at high temperature to assure that the equilibrium state was reached before being quenched to room temperature (300 K) with each specific cooling rate controlled by a Nose-Hoover thermostat (the volume of the sample was controlled through the use of a barostat set to zero pressure) [37]. Periodic boundary conditions were applied in all three directions. The Voronoi tessellation analysis was employed to investigate the short range order (SRO) according to nearest neighbor atoms from their inherent structures [7]. The inherent structure was obtained by conjugate-gradient energy minimization with energy threshold of $10^{-6}$ eV and force threshold of $10^{-6}$ eV/Å$^3$. The structure analysis of liquids were averaged over 100 configurations for each sample with running time of 1 ns.

**Elastic deformation of metallic glasses:** Athermal quasi-static shear (AQS) [54] was applied to the metallic glasses under study to avoid any strain rate effects and thermal fluctuations in the MD simulations; a simple shear was applied in the y-z direction in the nominal elastic regime with the strain range of zero to 0.05. To investigate the atomic cluster strain for four different cluster connection schemes, we modified a previous approach for atomic strain proposed by Falk [58] and Li [59], which was determined by minimizing the mean-square difference between previous and present configuration of atomic clusters. Here the first step was to seek a locally affine transformation matrix $J_{cluster,\alpha}$, which can be best used to  map:

$$\{\Delta R_{ji}^0\} \rightarrow \{\Delta R_{ji}\}; \ \forall i, j \in N_\alpha,$$

where $\Delta R_{ji}^0$ and $\Delta R_{ji}$ are the separation between central $i$ atom and surrounding $j$ atom at the second nearest-neighbor shell for previous and present configurations, respectively, while $\alpha$ refers to each cluster connection schemes of *1-atom* to *4-atom*. The Lagrangian strain matrix for $\alpha$ cluster connection schemes can then be calculated as:

$$\eta_k = \frac{1}{2}\left(J_{cluster,\alpha} \cdot J^T{}_{cluster,\alpha} - I\right),$$

such that its component in the *y-z* direction is the atomic cluster strain for a specific cluster connection.

**Reference:**

1. Na, J. H., Demetriou, M. D., Floyd, M., Hoff, A., Garrett, G. R., & Johnson, W. L. Compositional landscape for glass formation in metal alloys. *Proceedings of the National Academy of Sciences*, *111*(25), 9031-9036. (2014)

2. Inoue, A. Stabilization of metallic supercooled liquid and bulk amorphous alloys. *Acta Materialia 48(1)*:279-306. (2000)

3. Greer, A. L., & Ma, E. Bulk metallic glasses: at the cutting edge of metals research. *MRS bulletin*, *32*(08), 611-619. (2007)

4. Demetriou, M. D., *et al*. A damage-tolerant glass. *Nature Materials*, *10*(2), 123-128 (2011)

5. Stillinger, F. H. A topographic view of supercooled liquids and glass formation. *Science*, *267(5206)*, 1935-1939. (1995)

6. Debenedetti, P. G., & Stillinger, F. H. Supercooled liquids and the glass transition. *Nature*, *410(6825)*, 259-267 (2001)

7. Cheng, Y. Q. & Ma, E. Atomic-level structure and structure-property relationship in metallic glasses. *Progress in Materials Science 56(4):*379-473. (2011)

8. Ma, E. Tuning order in disorder, *Nature Materials*. *14*, 547-552 (2015)

9. Sheng, H. W., Luo, W. K., Alamgir, F. M., Bai, J. M., & Ma, E. Atomic packing and short-to-medium-range order in metallic glasses. *Nature 439(7075)*:419-425. (2006)

10. Miracle, D. B. A structural model for metallic glasses. *Nature Materials 3(10)*:697-702. (2004)

11. Egami, T. Atomic level stresses. *Progress in Materials Science 56(6)*:637-653. (2011)

12. Nelson, D. R. & Spaepen, F. Polytetrahedral order in condensed matter. *Solid State Physics 42*:1-90. (1989)

13. Shen, Y. T., Kim, T. H., Gangopadhyay, A. K., & Kelton, K. F. Icosahedral order, frustration, and the glass transition: Evidence from time-dependent nucleation and supercooled liquid structure studies. *Phys. Rev. Lett.*, *102*(5), 057801. (2009).

14. Hirata, A., *et al.* Direct observation of local atomic order in a metallic glass. *Nature Materials 10(1)*:28-33. (2011)

15. Hirata, A., *et al*. Geometric Frustration of Icosahedron in Metallic Glasses. *Science 341(6144)*:376-379. (2013)

16. Ding, J., Cheng, Y. Q., & Ma, E. Full icosahedra dominate local order in $Cu_{64}Zr_{34}$ metallic glass and supercooled liquid. *Acta Materialia 69*:343-354. (2014)

17. Ding, J., Cheng, Y. Q., Sheng, H. W., & Ma, E. Short-range structural signature of excess specific heat and fragility of metallic-glass-forming supercooled liquids. *Physical Review B 85(6).* (2012)

18. Ding, J., Cheng, Y. Q., & Ma, E. Quantitative measure of local solidity/liquidity in metallic glasses. *Acta Materialia 61(12)*:4474-4480. (2013)

19. Ding, J., Patinet, S., Falk, M. L., Cheng, Y. Q., & Ma, E. Soft spots and their structural signature in a metallic glass. *Proceedings of the National Academy of Sciences 111(39)*, 14052-14056. (2014)

20. Ma, D., Stoica, A. D., & Wang, X. L. Power-law scaling and fractal nature of medium-range order in metallic glasses. *Nature Materials 8(1):*30-34. (2009)

21. Zeng, Q., *et al.* Universal fractional noncubic power law for density of metallic glasses. *Physical Review Letters 112(18)*:185502. (2014)

22. Liu, X. J., *et al*. Metallic liquids and glasses: atomic order and global packing. *Physical Review Letters 105*(15):155501. (2010)

23. Häussler, P.. Analogies and differences between the crystalline and the disordered state. *Physica Status Solidi (c)*, *1*(*11*), 2879-2883. (2004)

24. Häussler, P. Interrelations between atomic and electronic structures—liquid and amorphous metals as model systems. *Physics Reports*, *222*(*2*), 65-143. (1992)

25. Fang, X. W., *et al.* Spatially resolved distribution function and the medium-range order in metallic liquid and glass. *Scientific Reports 1*, 194 (2011).

26. Li, M., Wang, C. Z., Hao, S. G., Kramer, M. J., & Ho, K. M.. Structural heterogeneity and medium-range order in ZrxCu100−x metallic glasses. *Physical Review B*, *80*(*18*), 184201. (2009).

27. Wu, Z. W., Li, M. Z., Wang, W. H., & Liu, K. X.. Hidden topological order and its correlation with glass-forming ability in metallic glasses. *Nature Communications 6, 6035* (2015).

28. Wakeda, M., and Shibutani. Y. Icosahedral clustering with medium-range order and local elastic properties of amorphous metals. *Acta Materialia* 58(11),3963-3969. (2010)

29. Wu, Z. W., Li, M. Z., Wang, W. H., & Liu, K. X. Correlation between structural relaxation and connectivity of icosahedral clusters in CuZr metallic glass-forming liquids. *Physical Review B*, *88*(5), 054202 (2013)

30. Zhang, Y., Zhang, F., Wang, C. Z., Mendelev, M. I., Kramer, M. J., & Ho, K. M. Cooling rates dependence of medium-range order development in Cu64.5Zr35.5 metallic glass. *Physical Review B*, *91*(6), 064105. (2015)

31. Bennett, C. H. Serially deposited amorphous aggregates of hard spheres. *Journal of applied physics*, *43*(6), 2727-2734. (1972).

32. Lee, M., Lee, C. M., Lee, K. R., Ma, E., & Lee, J. C. Networked interpenetrating connections of icosahedra: Effects on shear transformations in metallic glass. *Acta Materialia*, *59*(1), 159-170. (2011).

33. Pan, S. P., Qin, J. Y., Wang, W. M., & Gu, T. K. Origin of splitting of the second peak in the pair-distribution function for metallic glasses. *Physical Review B 84(9)*: 092201. (2011)

34. Cheng, Y. Q., Ma, E., & Sheng, H. W. Atomic level structure in multicomponent bulk metallic glass. *Physical Review Letters 102*(24). (2009)

35. Luo, W. K., Sheng, H. W., & Ma, E. Pair correlation functions and structural building schemes in amorphous alloys. *Applied Phyiscs Letters 89*, 131927 (2006)

36. van de Waal, B. W. On the origin of second-peak splitting in the static structure factor of metallic glasses. *Journal of non-crystalline solids*, *189*(1), 118-128. (1995)

37. Allen, M. P., Tildesley, D. J. Computer Simulation of Liquids (Clarendon Press, Oxford, UK). (1987)

38. Ding, J., Cheng, Y. Q., & Ma, E. Charge-transfer-enhanced prism-type local order in amorphous $Mg_{65}Cu_{25}Y_{10}$: Short-to-medium-range structural evolution underlying liquid fragility and heat capacity. *Acta Materialia 61(8)*:3130-3140. (2013)

39. Ding, J., & Cheng, Y. Q. Charge transfer and atomic-level pressure in metallic glasses. *Applied Physics Letters*, *104*(5), 051903. (2014).

40. Zhong, L., Wang, J., Sheng, H., Zhang, Z., & Mao, S. X. Formation of monatomic metallic glasses through ultrafast liquid quenching. *Nature*, *512* (7513), 177-180.(2014)

41. Stillinger, F. H., & Weber, T. A.Packing structures and transitions in liquids and solids. *Science*, *225*(4666), 983-9.(1984)

42. Jonsson, H., & Andersen, H. C. Icosahedral ordering in the Lennard-Jones liquid and glass. *Physical Review Letters*, *60*(22), 2295-2298. (1988)

43. Jakse, N., & Pasturel, A. Local order of liquid and supercooled zirconium by ab initio molecular dynamics. *Physical Review Letters*, *91*(19), 195501.(2003)

44. Stillinger, F. H., & Weber, T. A. Computer simulation of local order in condensed phases of silicon. *Physical review B*, *31*(8), 5262.(1985)

45. Ding, J., *et al*. Temperature effects on atomic pair distribution functions of melts. *The Journal of Chemical Physics*, *140*(6), 064501. (2014).

46. Smithells, C. J. Smithells Metals Reference Book, 7th ed., edited by E. A. Brandes and G. B. Brook. Butterworth-Heinemann, London (1992)

47. Liu, X. J., *et al.* (2011). Atomic packing symmetry in the metallic liquid and glass states. *Acta Materialia*, *59(16)*, 6480-6488. (2011).

48. Gardner, P. P., Cowlam, N., & Davies, H. A. Measurements of metallic glass structures under conditions of high spatial resolution. *Journal of Physics F: Metal Physics*, *15*(4), 769. (1985)

49. Fang, H. Z., Hui, X., Chen, G. L., & Liu, Z. K. Al-centered icosahedral ordering in Cu46Zr46Al8 bulk metallic glass. *Applied Physics Letters*, *94*(9), 1904.(2009)

50. Dmowski, W., Iwashita, T., Chuang, C-P., Almer, J., Egami, T. Elastic heterogeneity in metallic glasses. *Physical Review Letters. 105*, 205502 (2010)

51. Poulsen, H. F., Wert, J. A., Neuefeind, J., Honkimäki, V., & Daymond, M. Measuring strain distributions in amorphous materials. *Nature Materials*, *4*(1), 33-36. (2005).

52. Hufnagel, T. C., Ott, R. T., & Almer, J. Structural aspects of elastic deformation of a metallic glass. *Physical Review B*, *73*(6), 064204. (2006).

53. Ding, J., Cheng, Y. Q., & Ma, E. Correlating local structure with inhomogeneous elastic deformation in a metallic glass. *Applied Physics Letters* 101(12). (2012)

54. Maloney, C. E., & Lemaitre, A. Amorphous systems in athermal, quasistatic shear. *Physical Review E*, *74*(1), 016118. (2006).

55. Johnson, W. L., & Samwer, K. A universal criterion for plastic yielding of metallic glasses with a (T/T g) 2/3 temperature dependence. *Physical Review Letters*, *95*(19), 195501. (2005).

56. Johnson, W. L., Demetriou, M. D., Harmon, J. S., Lind, M. L., & Samwer, K. Rheology and ultrasonic properties of metallic glass-forming liquids: A potential energy landscape perspective. *MRS bulletin*, *32*(08), 644-650. (2007).

57. Plimpton, S. Fast parallel algorithms for short-range molecular dynamics. *Journal of computational physics*, *117*(1), 1-19. (1995)

58. Falk, M. L., & Langer, J. S. Dynamics of viscoplastic deformation in amorphous solids. *Physical Review E*, *57*(6), 7192 (1998).

59. Li, J. AtomEye: an efficient atomistic configuration viewer. *Modelling and Simulation in Materials Science and Engineering*, *11*(2), 173. (2003).

## Acknowledgements

This work was supported by the Mechanical Behavior of Materials Program at the Lawrence Berkeley National Laboratory, funded by the U.S. Department of Energy, Office of Science, Office of Basic Energy Sciences, Materials Sciences and Engineering Division, under Contract No. DE-AC02-05CH11231. E.M. has been supported at JHU by National Science Foundation, DMR-1505621. This work made use of resources of the National Energy Research Scientific Computing Center, supported by the Office of Basic Energy Sciences of the U.S. Department of Energy under Contract No. DE-AC02-05CH11231.

**Competing financial interests:** The authors declare no competing financial interests.

**Author Contribution:** J.D. conceived and designed the research and analysis, and performed the molecular dynamics simulations. J.D., E.M. M.A., R.R. wrote the paper. All authors discussed the results and revised the manuscript.

**Figure legends**

**Fig. 1.** (a) Radial distribution functions $g(r)$ of Ta liquids at 3300 K (orange line) and the inherent structures of Ta liquids at 3300K (cyan line) as well Ta glass at 300 K (blue dashed line). The inset schematically illustrates the atomic order at the second nearest-neighbor shell as the correlation between two central atoms (as marked), which also corresponds to the second peak in $g(r)$. (b) Shown schematically are four different schemes of coordination polyhedra connections with the number of shared atoms from one to four, which are denoted as *1-atom*, *2-atom*, *3-atom*, and *4-atom* cluster connections, respectively.

**Fig. 2.** Radial distribution functions $g(r)$ for (a) $Ni_{80}P_{20}$ and (b) $Zr_{46}Cu_{46}Al_8$ MGs at 300 K obtained by MD simulation with a cooling rate of $10^{10}$ K/s (Samples #5 and 8, respectively, in Table I). The decomposed radial distribution functions for nearest-neighbors (NN), second nearest neighbor atoms via *1-atom, 2-atom, 3-atom* and *4-atom* cluster connections, are also included. The insets show a magnified RDF at large distances.

**Fig. 3.** (a) The correlation between average bond length $R_1$ and average distance ($R_2^*$) of the second nearest-neighbors, for the four cluster connection schemes. Each data point is from each species in the samples listed in Table I. The solid lines are from a geometric calculation, as described in the text. (b) The average number of connected clusters versus NN coordination number for each species of the studied samples (in Table I) at their corresponding liquidus temperatures. We calculate $R_1$ and CN for each species, in each sample. For example, Cu in $Cu_{64}Zr_{36}$ and $Zr_{46}Cu_{46}Al_8$ have different measured $R_1$ and CN and they are all used in Fig. 3.

**Fig. 4.** (a) The number of connected clusters per atom, N, for four cluster connection schemes (*1-atom, 2-atom, 3-atom* and *4-atom*) at room temperature ($T$=300 K) for samples #1 to 8 in Table I. (b) The difference in N for each of the four cluster connection schemes (*1-atom, 2-atom, 3-atom* and *4-atom*) between T=300 K and the liquidus temperature for sample #1 to 8 in Table I. The first four samples as marked, are $Cu_{64}Zr_{36}$ MGs prepared at increasing cooling rates.

**Fig. 5.** The relation between cluster shear strains, $\gamma_{cluster}$, of different cluster connection schemes in terms of the imposed macroscopic shear strain $\gamma_{imposed}$ for samples #1 to 8 in Table I. The dashed line indicates where the cluster strains are equal to the imposed macroscopic shear strains.

**Table I.** Metallic glass samples prepared by MD simulation for analysis in this work.

| Sample # | Comp. | # of Atoms | Cooling rate (K/s) | Box length |
| --- | --- | --- | --- | --- |
| 1 | $Cu_{64}Zr_{36}$ | 128,000 | $10^9$* | 12.74 nm |
| 2 | $Cu_{64}Zr_{36}$ | 128,000 | $10^{10}$ | 12.75 nm |
| 3 | $Cu_{64}Zr_{36}$ | 128,000 | $10^{11}$ | 12.76 nm |
| 4 | $Cu_{64}Zr_{36}$ | 128,000 | $10^{12}$ | 12.77 nm |
| 5 | $Ni_{80}P_{20}$ | 128,000 | $10^{10}$ | 11.24 nm |
| 6 | $Al_{90}La_{10}$ | 128,000 | $10^{10}$ | 13.44 nm |
| 7 | $Mg_{65}Cu_{25}Y_{10}$ | 128,000 | $10^{10}$ | 13.89 nm |
| 8 | $Zr_{46}Cu_{46}Al_8$ | 128,000 | $10^{10}$ | 13.05 nm |

* Note that the quenching procedure for sample #1 was as follows: First, it was quenched to 1200 K with the cooling rate of $10^{10}$ K/s from equilibrium liquid at 2500K. It was then cooled to 600K (well below glass transition temperature of ~750K) at $10^9$ K/s, followed by quenching to room temperature at a cooling rate of $10^{10}$ K/s. Since the configurational state of the glass is mainly determined by the cooling rate within the supercooled region, the effective cooling rate of the sample can be regarded as approximately $10^9$ K/s.

Figure 1

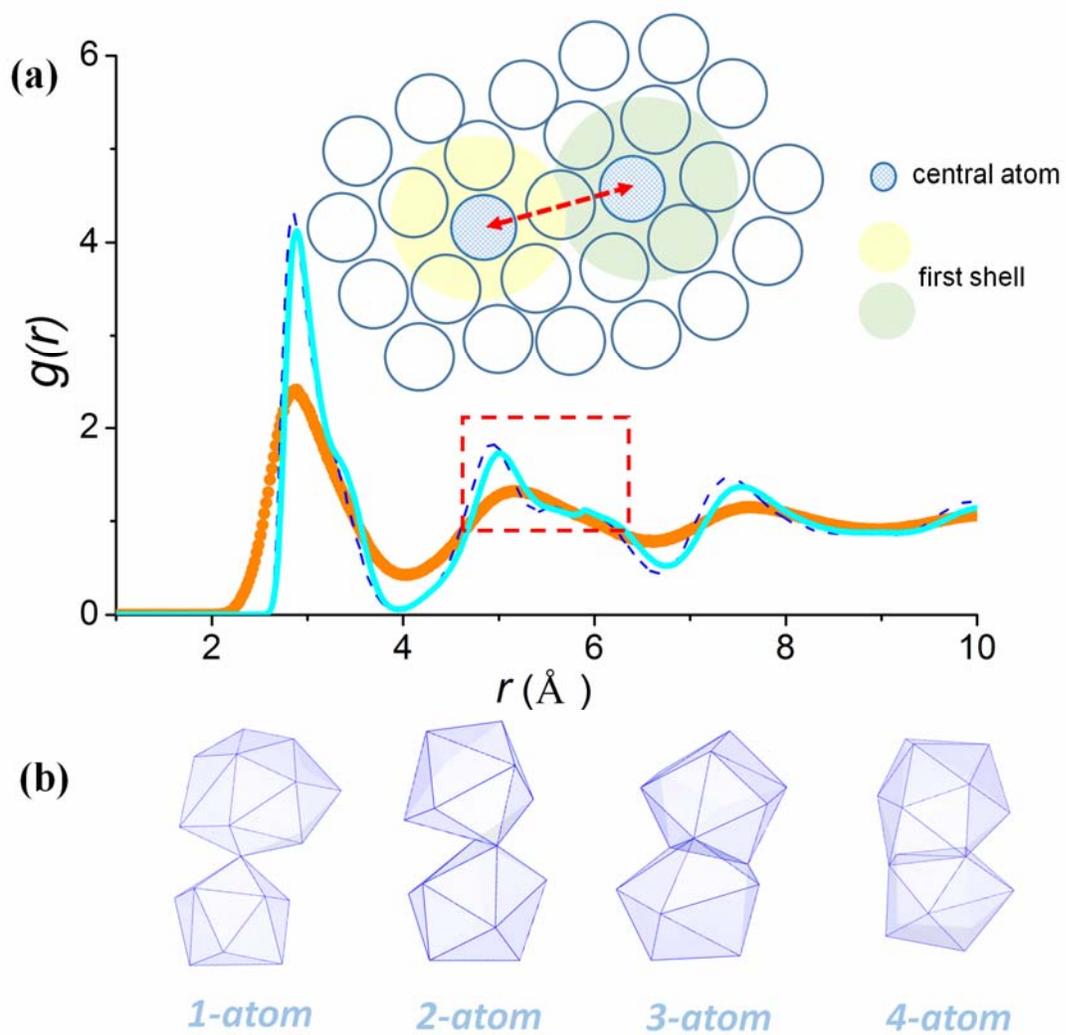

(a)

(b) 1-atom  2-atom  3-atom  4-atom

Figure 2

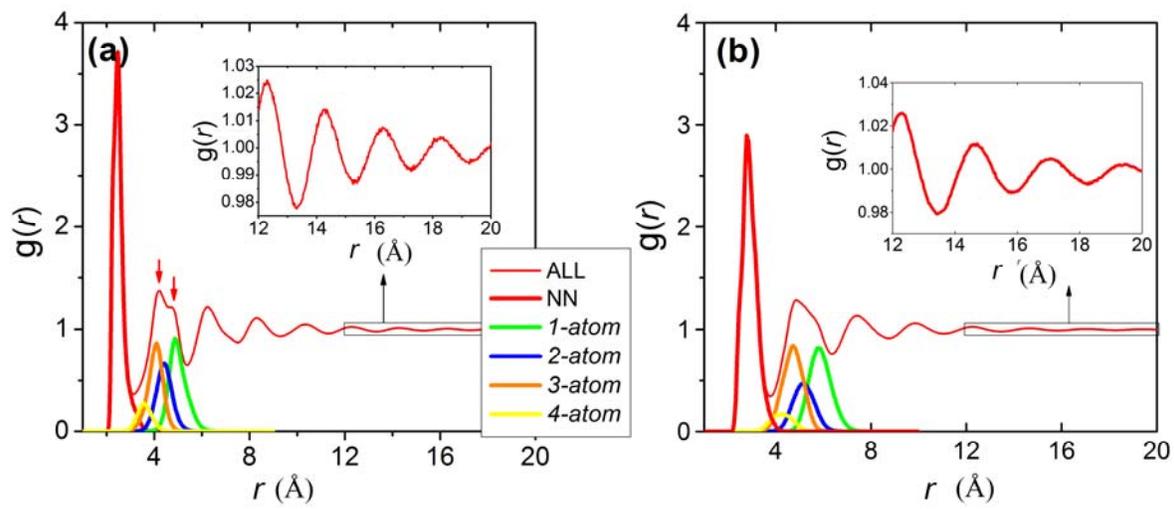

Figure 3

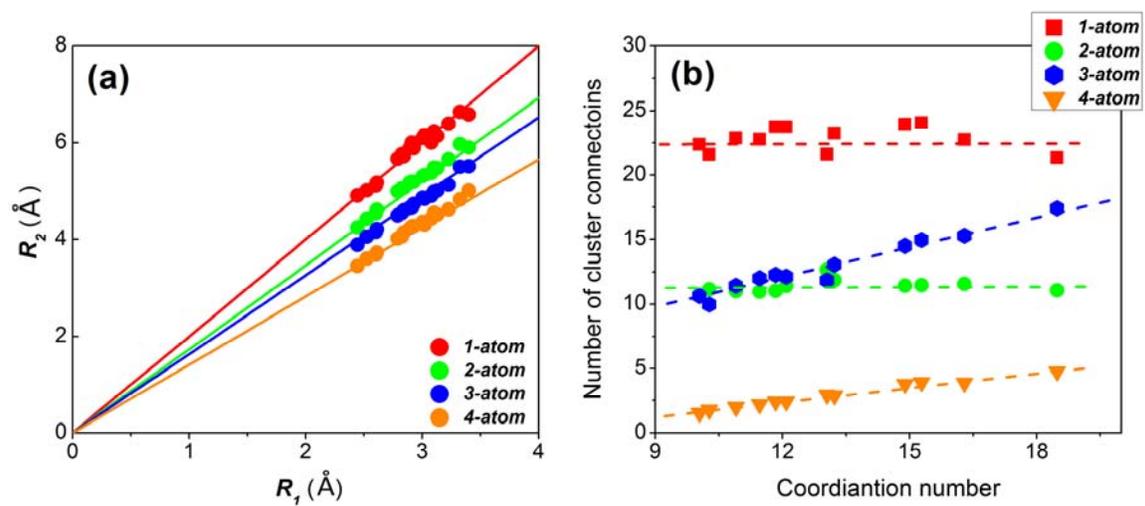

Figure 4

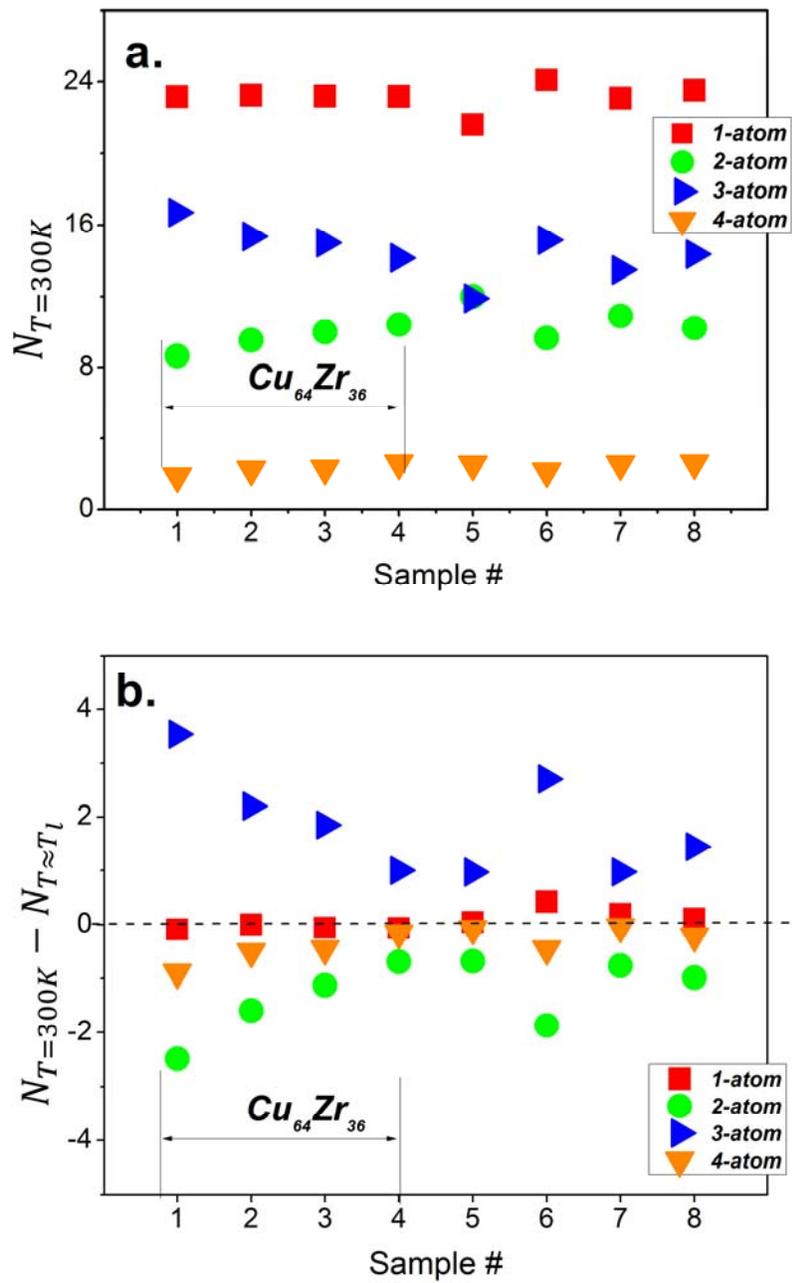

Figure 5

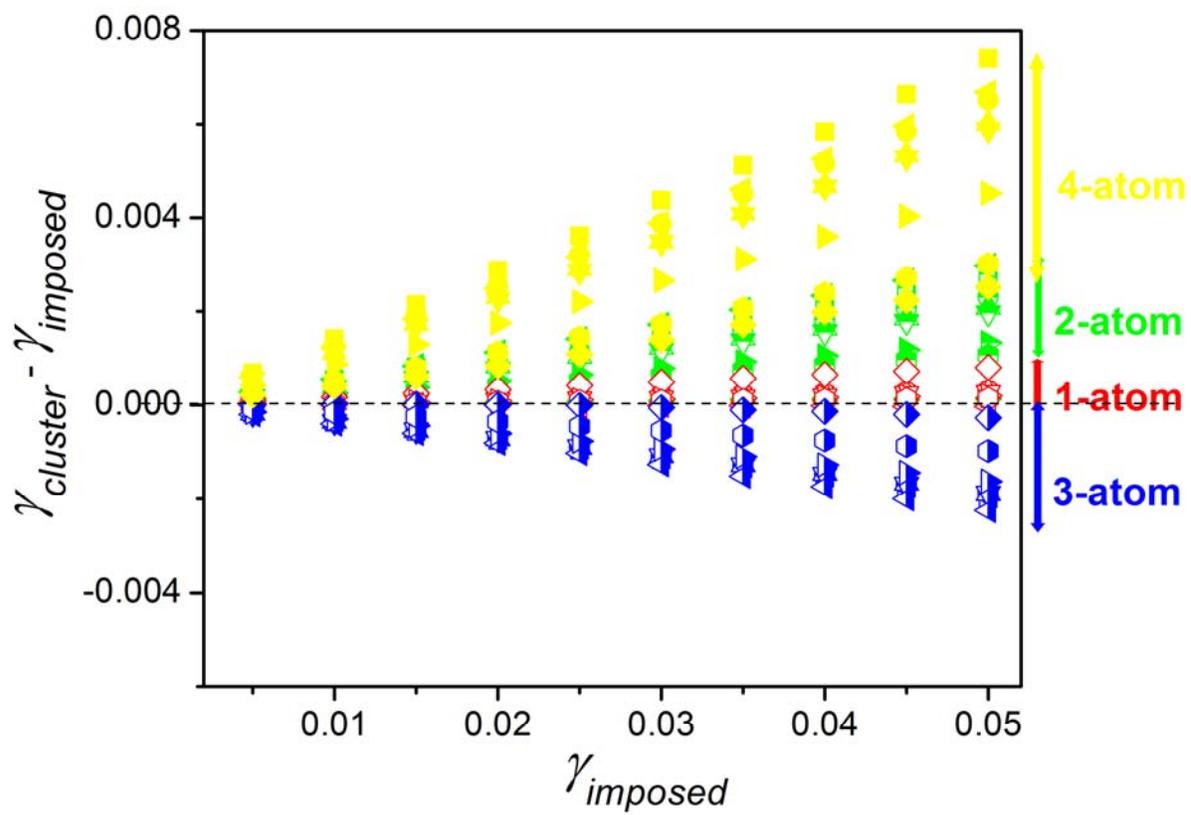